\documentclass[preprint]{elsarticle}
\usepackage{amsmath}
\usepackage{amssymb}
\usepackage{amsthm}
\usepackage{psfrag}
\usepackage{graphicx}
\usepackage{color}
\usepackage{upgreek}
\journal{Physics Letters B}
\begin{document}
\begin{frontmatter}
\title{Exact solutions of the Einstein equations for  an  infinite slab with a constant energy density. }

\author{Alexander Yu. Kamenshchik}
\address{Dipartimento di Fisica e Astronomia, Universit\`a di Bologna and INFN, Via Irnerio 46, 40126 Bologna,
Italy\\Physics Letters B 791, 201 (2019).
L.D. Landau Institute for Theoretical Physics of the Russian
Academy of Sciences, Kosygin str. 2, 119334 Moscow, Russia}\ead
{kamenshchik@bo.infn.it}
\author{Tereza  Vardanyan}
\address{Dipartimento di Fisica e Astronomia, Universit\`a di Bologna and INFN, Via Irnerio 46,  40126 Bologna,
Italy}\ead{tereza.vardanyan@bo.infn.it}

\begin{abstract}
We find exact static solutions of the Einstein equations in the spacetime with plane symmetry, where an infinite slab with finite thickness
and homogeneous energy (mass) density is present. In the first solution the pressure is isotropic, while in the second solution the tangential 
components of the pressure are equal to zero. In both cases the pressure vanishes at the boundaries of the slab. Outside  the slab these solutions are matched with the Rindler
spacetime and with the Weyl-Levi-Civita spacetime, which represent special cases of the Kasner solution.  
\end{abstract}
 
\end{frontmatter}

\section{Introduction}
It is known that even in the absence of matter sources the Einstein equations of General relativity can have very nontrivial solutions. 
Historically the first such solution was the external Schwarzschild solution for a static spherically symmetric geometry \cite{Schwarz}.
It was extremely  useful for study of general relativistic corrections to the Newtonian gravity and for the description of such effects as the precession 
of the Mercury perihelion and the light deflection in the gravitational field. This solution also opened a fruitful field of black hole physics.
The Schwarzschild solution contains a genuine singularity in the centre of the spherical symmetry. To avoid it and to describe real spherically symmetric objects like stars, Schwarzschild also invented an internal solution \cite{Schwarz1} generated by a ball with constant energy density and with isotropic pressure. At the boundary of the ball the pressure  disappears and the external and internal solutions are matched.   In this case there is no singularity in the center of the ball. Later, more general spherically symmetric geometries were studied in the papers by Tolman \cite{Tolman}, Oppenheimer-Volkoff \cite{Op-Vol}, Buchdahl  \cite{Buchdahl} and many others. Similar problems with cylindrical axial symmetry were also studied (see, e.g. \cite{Fulling0} and references therein). In paper 
\cite{Geroch} the question of existence of  solutions of the Einstein equations in the presence of concentrated matter sources, described by the generalised functions 
(distributions) was studied. It was shown, that in contrast to the case of electrodynamics, where the charged ball can be contracted to the point and the charge density 
becomes proportional to the Dirac delta function while the Poisson equation is still valid, we cannot do it in the General Relativity. The reason lies in the non-linearity of the Einstein equations. It was shown in \cite{Geroch} that the solutions with distributional sources cannot exist for zero-dimensional (point-like particles) and one-dimensional 
(strings) objects, but can exist for two-dimensional (shells) objects. This fact makes the study of geometries possessing plane symmetries particularly interesting.
Indeed, the plane-symmetric solutions of the Einstein equations were also studied in literature (see e.g. \cite{Amundsen, Fulling} and the references therein). 

However, to our knowledge, exact static solutions of the Einstein equations, in the spacetimes with plane symmetry in the presence of an infinite slab with 
a finite thickness were not studied. Thus, our objective in the present paper was to find such solutions with the matching between the geometry inside the slab and that outside of it.  Here, we would like to say that the first static solution in an empty spacetime possessing  plane symmetry is almost as old as the Schwarzschild solution.  
This is the spatial Kasner solution \cite{Kasner} found in 1921 and its particular case - the Weyl-Levi-Civita solution \cite{Weyl,Levi-Civita}, found even earlier.  
Hence, we wanted to find for the case of plane symmetry some analog of matching between Schwarzschild external and internal solutions. We have considered 
an infinite slab with a finite thickness and a constant mass (energy) density and have found two particular solutions: one with isotropic pressure and one for tangential pressure equal to zero. In both cases we require that all components of pressure vanish at the boundary of the slab, just like in the case of the Schwarzschild internal solution.  The structure of the paper is the following: in the second section we write down some general formulae  for the spacetimes with the spatial geometry possessing plane symmetry;
in section 3 we describe the solution with  isotropic pressure, while the section 4 is devoted to the solutions with  vanishing tangential pressure. The fifth section contains some concluding remarks.        

\section{Einstein equations for spacetimes with spatial geometry possessing plane symmetry}
Let us consider the metric with plane symmetry, where the metric coefficients depend on one spatial coordinate $x$:
\begin{equation}
ds^2=a^2(x)dt^2-dx^2-b^2(x)dy^2-c^2(x)dz^2.
\label{metric}
\end{equation}
Before plunging into technical details connected with the search for the solutions for the thick slab, let us recall briefly what is known about the empty spacetime solutions and the solutions in the presence of thin shells.

For the metric (\ref{metric}) in the empty spacetime we have two general solutions. One of them is the Minkowski metric,
where $a=b=c=1$ and another one is the Kasner solution \cite{Kasner} with 
\begin{equation}
a(x) = a_0(x-x_1)^{p_1},\ b(x) = b_0(x-x_1)^{p_2},\ c(x) =   c_0(x-x_1)^{p_3},
\label{Kasner}
\end{equation}
where the Kasner indices $p_1,p_2$ and $p_3$ satisfy the equations
\begin{equation}
p_1+p_2+p_3=p_1^2+p_2^2+p_3^2 = 1.
\label{Kasner1}
\end{equation}
The Kasner solution is more often used in a ``cosmological form'':
\begin{equation}
ds^2=dt^2-a_0^2t^{2p_1}dx^2-b_0^2t^{2p_2}dy^2-c_0^2t^{2p_3}.
\label{Kasner2}
\end{equation}
This form of the Kasner metric was rediscovered in papers \cite{Taub,Heck-Schuck,Khal-Lif} and has played an important 
role in cosmology. The study of Kasner dynamics in  paper \cite{Khal-Lif} has led to the discovery of the oscillatory approach to the cosmological singularity \cite{BKL}, known also as the Mixmaster universe \cite{Misner}. The further development of this line of research has brought the establishment of the connection between the chaotic behaviour of the universe in superstring  models  
 and the infinite-dimensional Lie algebras \cite{Damour}.
 
 Coming back to the spatial form of the Kasner metric (\ref{Kasner})-(\ref{Kasner1}), one sees that the requirement of  symmetry in the plane between the  $y$ and $z$ 
 directions implies the condition
 \begin{equation}
 p_2=p_3.
 \label{Kasner3}
 \end{equation}
 It is easy to see that there are two solutions of Eqs. (\ref{Kasner1}) satisfying the condition (\ref{Kasner3}). One of them 
 is the Rindler spacetime \cite{Rindler} 
 with 
 \begin{equation}
 p_1=1,\ p_2 = p_3 =0.
 \label{Rindler}
 \end{equation}
 It is well-known that the Rindler spacetime represents a part of the Minkowski spacetime rewritten in the coordinates connected 
 with an accelerated observer. There is a coordinate singularity (horizon) at $x=x_1$.
 Another solution is 
 \begin{equation}
 p_1=-\frac13,\ p_2=p_3=\frac23.
 \label{Weyl}
 \end{equation} 
 This particular solution was found by Weyl \cite{Weyl} and Levi-Civita \cite{Levi-Civita} before the work of Kasner\footnote{In paper \cite{Khal-Lif} a convenient parametrization of the Kasner indices was presented: 
 \begin{equation}
  p_1=-\frac{u}{1+u+u^2},\ p_2 = \frac{1+u}{1+u+u^2},\ p_3 = \frac{u(1+u)}{1+u+u^2}.
  \nonumber
  \end{equation}
  In terms of this parametrization, the Rindler solution corresponds to $u=0$, while the Weyl-Levi-Civita solution is given by 
  $u=1$. 
}. This solution describes a universe, where a real curvature singularity is present at $x=x_1$.

The detailed account of the solutions in the presence of  a thin plate with constant energy density was given in paper \cite{Fulling}. These solutions have some distinguishing features.
First of all the energy density of the plate and its tangential pressure    should  both be proportional to the delta function, while the component of the pressure, perpendicular to the plate is equal to zero. Furthermore, the metric is continuous everywhere, but its derivative has a finite jump at the location of the plate.   
The spacetimes on the right and on the left from the plane are of the type described above: Minkowski, Rindler or Weyl-Levi-Civita. 
 The reflection symmetry is present, i.e. the spacetimes on both sides are the same, if and only if the energy density and the pressure are connected by the relation $p=-\frac14 \rho$ or $\rho=0$.  Otherwise  this symmetry is lost. 

In the paper \cite{Fulling} the solutions in the presence of a finite-thickness slab were also discussed. Some features of such solutions 
were analysed qualitatively or numerically, but exact solutions were not found. One of these interesting features is the absence of the reflection symmetry. Here we obtain some exact solutions manifesting this feature. Concerning the properties of the matter constituting the slab,  being inspired by the internal Schwarzschild solution \cite{Schwarz1}, we assume that the  energy density is constant while the pressure should disappear at the boundaries of the slab.    

Now we write down some general formulae necessary for the metric with the plane symmetry (\ref{metric}).
The non-vanishing Christoffel symbols are 
\begin{eqnarray}
&&\Gamma_{tt}^x=a'a,\ \Gamma_{yy}^x=-b'b,\ \Gamma_{zz}^x=-c'c,\nonumber \\
&&\Gamma_{tx}^t=\frac{a'}{a},\ \Gamma_{yx}^y=\frac{b'}{b},\ \Gamma_{zx}^z=\frac{c'}{c}.
\label{Christoffel}
\end{eqnarray}
The components of the Ricci tensor are
\begin{eqnarray}
&&R_{tt}=a''a+\frac{a'b'a}{b}+\frac{a'c'a}{c},\nonumber \\
&&R_t^t=\frac{a''}{a}+\frac{a'b'}{ab}+\frac{a'c'}{ac},\nonumber\\
&&R_{xx}=-\frac{a''}{a}-\frac{b''}{b}-\frac{c''}{c},\nonumber \\
&&R_x^x=\frac{a''}{a}+\frac{b''}{b}+\frac{c''}{c},\nonumber \\
&&R_{yy}=-b''b-\frac{a'b'b}{a}-\frac{b'c'b}{c},\nonumber \\
&&R_y^y=+\frac{b''}{b}+\frac{a'b'}{ab}+\frac{b'c'}{bc},\nonumber \\
&&R_{zz}=-c''c-\frac{a'c'c}{a}-\frac{b'c'c}{b},\nonumber \\
&&R_z^z=\frac{c''}{c}+\frac{a'c'}{ac}+\frac{b'c'}{bc}.
\label{Ricci}
\end{eqnarray}
The Ricci scalar is
\begin{equation}
R=2\left(\frac{a''}{a}+\frac{b''}{b}+\frac{c''}{c}+\frac{a'b'}{ab}+\frac{a'c'}{ac}+\frac{b'c'}{bc}\right).
\label{Ricci-scal}
\end{equation}
The energy-momentum tensor for a fluid with isotropic pressure   is 
\begin{equation}
T_{\mu\nu} = (\rho+p(x))u_{\mu}u_{\nu}-p(x)g_{\mu\nu},
\label{energy}
\end{equation}
where we shall write 
\begin{equation}
\rho = \frac{4k^2}{3} = \rm constant
\label{energy1}
\end{equation}
for convenience.
Then
\begin{equation}
u_t=a,\ u_x=u_y=u_z=0.
\label{velocity}
\end{equation}
The equation
\begin{equation}
T_{\mu;\nu}^{\nu} = 0
\label{pressure}
\end{equation}
for $\mu= x$ gives 
\begin{equation}
p'=-\frac{a'}{a}(\rho+p),
\label{pressure1}
\end{equation}
where  ``prime'' means the derivative with respect to $x$. 
The integration of Eq. (\ref{pressure1}) gives
\begin{equation}
p = -\frac{4k^2}{3} +\frac{p_0}{a},
\label{pressure2}
\end{equation}
where $p_0$ is an arbitrary constant.
The Einstein equations are
\begin{equation}
-\frac{b''}{b}-\frac{c''}{c}-\frac{b'c'}{bc}=\frac{4k^2}{3},
\label{Ein1}
\end{equation}
\begin{equation}
\frac{a'b'}{ab}+\frac{a'c'}{ac}+\frac{b'c'}{bc}=p,
\label{Ein2}
\end{equation}
\begin{equation}
+\frac{a''}{a}+\frac{c''}{c}+\frac{a'c'}{ac}=p,
\label{Ein3}
\end{equation}
\begin{equation}
+\frac{a''}{a}+\frac{b''}{b}+\frac{a'b'}{ab}=p.
\label{Ein4}
\end{equation}
Introducing new functions 
\begin{equation}
A = \frac{a'}{a},\ B=\frac{b'}{b},\ C = \frac{c'}{c},
\label{new}
\end{equation}
we can rewrite the Einstein equations (\ref{Ein1})-(\ref{Ein4}) as follows:
\begin{equation}
-B'-B^2-C'-C^2-BC = \frac{4k^2}{3},
\label{Ein1n}
\end{equation}
\begin{equation}
AB+AC+BC=p,
\label{Ein2n}
\end{equation}
\begin{equation}
A'+A^2+C'+C^2+AC = p,
\label{Ein3n}
\end{equation}
\begin{equation}
A'+A^2+B'+B^2+AB=p.
\label{Ein4n}
\end{equation}

\section{Solution with isotropic pressure}
In what follows we shall consider only the solutions where the symmetry between the directions along the coordinate axes $y$ and $z$ is present. Then  
\begin{equation}
B=C,
\label{isot}
\end{equation} 
and 
we obtain from Eq. (\ref{Ein1n})
\begin{equation}
-2B'-3B^2=\frac{4k^2}{3}.
\label{Ein1n1}
\end{equation}
Integrating this equation, we obtain 
\begin{equation}
B = C=-\frac23k\tan k(x+x_0).
\label{B}
\end{equation}
Using the definitions (\ref{new}), we obtain
\begin{equation}
b=b_0(\cos k(x+x_0))^{\frac23},
\label{b}
\end{equation}
\begin{equation}
c=c_0(\cos k(x+x_0))^{\frac23}.
\label{c}
\end{equation}
Let us note that in order to not have singularities in the metric, we need to require that 
\begin{equation}
 [-L+x_0,L+x_0]\subset (-\pi/2,\pi/2),
\label{range0}
\end{equation}  
where $x=-L$ and $x=L$ are the locations of the boundary of the slab.
Substituting Eqs. (\ref{B}) and (\ref{pressure2}) into Eq. (\ref{Ein2n}), we obtain
\begin{equation}
-\frac{a'}{a}\frac{4k}{3}\tan k(x+x_0)+\frac{4k^2}{9}\tan^2k(x+x_0)=- \frac{4k^2}{3}+\frac{p_0}{a}.
\label{a}
\end{equation}
This equation can be rewritten as 
\begin{eqnarray}
&&a'-\frac{k}{3}\tan k(x+x_0) a-k\cot k(x+x_0) a+\frac{3p_0}{4k}\cot k(x+x_0)=0.
\label{a1}
\end{eqnarray}
The general solution of the corresponding homogeneous equation is 
\begin{equation}
a(x) = a_1\sin k(x+x_0)(\cos k(x+x_0))^{-\frac13},
\label{homogen}
\end{equation} 
where $a_1$ is an integration constant. We shall look for  the
solution of the inhomogeneous equation (\ref{a1}) in the following form 
\begin{equation}
a(x) =\tilde{a}(x)\sin k(x+x_0)(\cos k(x+x_0))^{-\frac13}.
\label{tilde}
\end{equation}
Substituting the expression (\ref{tilde}) into Eq. (\ref{a1}), we have 
\begin{equation}
\tilde{a}'= -\frac{3p_0}{4k}\frac{(\cos k(x+x_0))^{\frac43}}{\sin^2k(x+x_0)}.
\label{tilde1}
\end{equation}  
Integrating by parts, we obtain
\begin{eqnarray}
&&\tilde{a}(x) =\frac{3p_0}{4k^2}\cot k(x+x_0)(\cos k(x+x_0))^{\frac43}\nonumber \\
&&+
\frac{p_0}{k}\int dx \left(\cos k(x+x_0)\right)^{\frac43} +a_2,
\label{tilde2}
\end{eqnarray}
 where $a_2$ is an integration constant. 
 Introducing a variable 
 $$
 u\equiv \sin^2 k(x+x_0)
 $$
 one can find that   
 \begin{eqnarray}
\frac{p_0}{k}\int_{-x_0}^x dy(\cos k(y+x_0))^{\frac43}=\frac{p_0}{2k^2} {\cal B}\left(\sin^2 k(x+x_0);\frac12,\frac76\right){\rm Sign} [\sin k(x+x_0)], 
\nonumber\\
\label{beta-Euler}
\end{eqnarray}
where the incomplete Euler function is defined as 
 \begin{equation}
 {\cal B}(x,r,s) \equiv \int_0^x du u^{r-1}(1-u)^{s-1}.
 \label{beta-Euler1}
 \end{equation}
 Thus, the general solution of Eq. (\ref{a1}) is 
 \begin{eqnarray}
 &&a(x) = \frac{3p_0}{4k^2}\cos^2 k(x+x_0)\nonumber \\
 &&+\frac{p_0}{2k^2}(\cos k(x+x_0))^{\frac13}|\sin k(x+x_0)|{\cal B}\left(\sin^2 k(x+x_0);\frac12,\frac76\right)\nonumber \\
&&+a_3\sin k(x+x_0)(\cos k(x+x_0))^{-\frac13}.
 \label{gen-sol}
 \end{eqnarray}
Looking at the expression (\ref{pressure2}), we see that the disappearance of the pressure on the boundary of the slab is equivalent to the requirement that 
\begin{equation}
a(-L) = a(L) = \frac{3p_0}{4k^2}.
\label{pressure3}
\end{equation}
On using Eq. (\ref{gen-sol}) this condition can be rewritten as 
\begin{eqnarray}
 && -\frac{3p_0}{4k^2}\sin^2k(\pm L+x_0)\nonumber \\
 &&+\frac{p_0}{2k^2}(\cos k(\pm L+x_0))^{\frac13}|\sin k(\pm L+x_0)|{\cal B}\left(\sin^2 k(\pm L+x_0);\frac12,\frac76\right)\nonumber \\
&&+a_3k\sin(\pm L+x_0)(\cos k(\pm L+x_0))^{-\frac13}=0.
 \label{gen-sol1}
 \end{eqnarray}
 Now, we have two free parameters $x_0$ and $a_3$, which we can fix in such a way to provide the disappearance of the pressure on the border of the slab. Let us first  choose 
 \begin{equation}
 x_0 = L.
 \label{x0}
 \end{equation}
 It guarantees that 
 \begin{equation}
 a(-L) = \frac{3p_0}{4k^2}
 \label{pressure4}
 \end{equation}
 and, hence, 
 \begin{equation}
 p(-L) = 0.
 \label{pressure5}
 \end{equation} 
With this choice of $x_0$ the requirement (\ref {range0}) becomes
 \begin{equation}
 2kL < \frac{\pi}{2}.
 \label{range}
 \end{equation}  
It is easy to see, that if the inequality (\ref{range}) is broken, the  cosine is equal to zero 
at some value of the coordinate $x$ inside the slab and one encounters a singularity.

 Now, substituting the value (\ref{x0}) into Eq. (\ref{gen-sol1}), we can choose the constant $a_3$ requiring the disappearance of
 the pressure on the other border of the slab $x=L$. This constant is 
\begin{equation}
a_3=\frac{p_0}{4k^2}\left(3\sin 2kL \cos^{1/3} 2kL-2\cos^{2/3} 2kL\ {\cal B}(\sin^2 2kL;1/2,7/6)\right).      
\label{a3}
\end{equation}    
 Finally we can write
 \begin{eqnarray}
 &&a(x) = \frac{3p_0}{4k^2}\cos^2k(x+L)\nonumber \\
 &&+\frac{p_0}{2k^2}(\cos k(x+L))^{\frac13}\sin k(x+L){\cal B}\left(\sin^2 k(x+L);\frac12,\frac76\right)\nonumber \\
&&+\frac{p_0}{4k^2}\left(3\sin 2kL \cos^{1/3} 2kL-2\cos^{2/3} 2kL\ {\cal B}
(\sin^2 2kL;1/2,7/6)\right)\nonumber \\
&&\times\sin k(x+L)(\cos k(x+L))^{-\frac13}.
 \label{gen-sol2}
 \end{eqnarray}    
 Thus, we have obtained a complete solution of the Einstein equations in the slab, where the energy density is constant and the pressure disappears on the boundary between the slab and an empty space.
Let us make  some comments here. First, the scale factors $a, b$ and $c$ and hence the metric coefficients are not even and the solution is not invariant with respect to the inversion 
$$
x \rightarrow -x.
$$ 
 However, making the change $x \rightarrow -x$ we obtain another solution of our equations. It can be obtained also by choosing $x_0 = -L$ instead of $x_0=L$ and by the corresponding change of the expression for the coefficient $a_3$, which is reduced to the change of the sign of the argument of the trigonometrical functions.  There is no qualitative difference between these two solutions. Thus, we shall study the first one. Let us emphasise  that the choice $x_0 = \pm L$ is  obligatory in order for the pressure to 
 vanish on  both boundaries of the slab  and, hence, the  asymmetry of these two solutions is an essential feature of the problem. It arises in spite of the initial symmetry of the Einstein equations and of the position of the slab. Thus, one can speak about  some kind of  symmetry breaking phenomenon.  
  
Let us consider the question of matching of the solutions in the slab with the vacuum solutions outside the slab. 
 Our solution inside the slab possesses  symmetry in the plane $(y,z)$. Thus, we shall try to match it at $x < -L$ and at $x > L$ with one of these three solutions: Minkowski, Rindler or Weyl-Levi-Civita (\ref{Weyl}). 
 
 Consider the plane $x=-L$. We shall require that 
\begin{eqnarray}
&&a_{\rm ext}(-L) = a(-L),\ b_{\rm ext}(-L) = b(-L),\ c_{\rm ext}(-L) = c(-L),\nonumber \\
&&a'_{\rm ext}(-L) = a'(-L),\ b'_{\rm ext}(-L) = b'(-L),\ c'_{\rm ext}(-L) = c'(-L).
\label{boundary}
\end{eqnarray}
Looking at the expressions (\ref{gen-sol2}), (\ref{b}), (\ref{c}) we see that at $x=-L$ the derivatives of $b$ and $c$ disappear 
(provided $x_0=L$), while the derivative of $a$ at this point is different from zero. Thus, we should choose the Rindler geometry for $x < -L$
\begin{equation}
ds^2=a_R^2(x-x_R)^2dt^2-dx^2-b_R^2(dy^2+dz^2).
\label{R}
\end{equation}
We can consider the analogous matching conditions at $x=L$. Here the derivatives of all three scale factors are non-vanishing. Thus, for $x > L$ we have a Weyl-Levi-Civita solution  
\begin{equation}
ds^2=a_{WLC}^2(x-x_{WLC})^{-2/3}dt^2-dx^2-b_{WLC}^2(x-x_{WLC})^{4/3}(dy^2+dz^2).
\label{WLC}
\end{equation}

Let us discuss now these matching conditions in more  detail. On the plane $x = -L$, we have 
\begin{equation}
\frac{3p_0}{4k^2}=a_R(-L-x_R),
\label{boundary1}
\end{equation}
to match the scale factors (the subscript ``R'' means ``Rindler'') and 
\begin{equation}
a_3 k = a_R,
\label{boundary2}
\end{equation}
where $a_3$ is given by Eq. (\ref{a3}) to match their derivatives. It follows from Eqs. (\ref{boundary1}) and (\ref{boundary2}) that 
\begin{equation}
x_R = -L - \frac{3p_0}{4a_3k^3}.
\label{boundary3}
\end{equation}
Plotting (\ref {a3}) as a function of $2kL$, we can see that for values smaller than $2kL\approx 1.05$ $a_3<0$ and thus
\begin{equation}
x_R >-L.
\label{boundary4}
\end{equation}
Therefore there is no horizon for these values of $kL$.

At the boundary $x=L$ it is more convenient to write down the conditions of matching of the tangential scale factors $b$:
\begin{equation}
b_0(\cos 2kL)^{2/3} = b_{WLC}(L-x_{WLC})^{2/3},
\label{boundary5}
\end{equation}
\begin{equation}
-\frac23b_0k(\cos 2kL)^{-1/3}\sin 2kL = \frac23b_{WLC}(L-x_{WLC})^{-1/3}.
\label{boundary6}
\end{equation}
From these two equations we easily find that 
\begin{equation}
x_{WLC} = L +\frac1k \cot 2kL.
\label{boundary7}
\end{equation}
Provided the condition (\ref {range}) we see that $x_{WLC}$ is necessarily bigger than $L$ and we can't avoid having a singularity in the space on the right side of the slab, at least not if the energy density $\rho$ of the slab is positive. To obtain the solution for the case $\rho<0$, we can replace $k$ by $ik$ in the solution that we already have. Then trigonometric functions turn into hyperbolic ones and the expression (\ref {boundary7}) is replaced by 
\begin{equation}
x_{WLC} = L -\frac1k \coth 2kL.
\label{neg}
\end{equation}
The above expression is smaller than $L$; therefore, there is no singularity. In the case of an infinitely thin slab, the conclusion that the singularity is unavoidable for $\rho>0$ was obtained in \cite{Fulling}.    

The expression for $x_{WLC}$ given by Eq. (\ref{boundary7})
guarantees the satisfaction of the matching conditions also for the scale factor $a$ and its derivative. It follows from the fact that for both the Weyl-Levi-Civita solution and 
for our internal solution 
\begin{equation}
\frac{a'}{a}(L) = -\frac12\frac{b'}{b}(L),
\label{boundary8}
\end{equation}
which in turn follows from Eq. (\ref{Ein2n}) and from the disappearance of the pressure on the border of the slab.     
 
As we mentioned earlier the solution (\ref {gen-sol2}) is not invariant with respect to the inversion of the coordinate $x$. However, for a particular value of $kL$ one can have an even solution, invariant with respect to this inversion. Indeed, we can transform the general solution  for the scale factor $a$  (\ref{gen-sol}) into an even function of $x$ by putting $a_3=0$ and $x_0=0$. Then also $b(x)$ and $c(x)$ become even. One can check numerically that  at $kL \approx 1.05$ the expression
 for $a$ at the boundaries $x=\pm L$ is such that the pressure disappears. The argument of the trigonometric functions runs between $-1.05 > -\frac{\pi}{2}$ and $1.05 < \frac{\pi}{2}$, the cosine is always different from zero and the singularity does not arise. Besides, at both boundaries the derivatives of the scale factors are different from zero. Hence, in both half-spaces outside the slab this solution should be matched with the Weyl-Levi-Civita solutions . Let us stress once again that this symmetric solution is a very particular one, arising at some special value of $kL$, while generally we have a pair of solutions, each of which is not symmetric with respect to the reflection $x \rightarrow -x$, instead  this reflection transforms one solution into another and vice versa.  One can trace here an analogy with a well-known case of two-well potential, which is often considered at the introducing of the spontaneous symmetry breaking phenomenon in quantum field theory (see e.g. \cite{Shirkov})
 $$
 V(\phi) = (\phi^2-\phi_0^2)^2,
 $$
 which is symmetric with respect to $\phi \rightarrow -\phi$, while its minimum values $\phi=\pm \phi_0$ are not symmetric.
 

\section{Solution with vanishing tangential pressure}
In the preceding section we have considered  a situation where the tangential pressure coincides with the transversal pressure, just like in the internal Schwarzschild solution \cite{Schwarz1}. In the case of the Schwarzschild spherically symmetric geometry, such a choice is obligatory because otherwise the pressure becomes infinite in the center of the ball and a non-singular internal solution does not exist (unless it is assumed that radial pressure is identically zero and tangential pressure does not vanish at the boundary; see \cite {Florides}). However, it is not obvious that in the case of the plane symmetry the situation is the same. Let us consider a more general energy-momentum tensor
\begin{equation}
T_t^t = \rho,\ T_x^x = -p_x,\ T_y^y = -p_y,\ T_z^z = -p_z. 
\label{en-mom-gen}
\end{equation}
Then the energy-momentum tensor conservation condition (\ref{pressure}) takes the following form
\begin{equation}
p_x'+A(\rho+p_x)+B(p_x-p_y)+C(p_x-p_z)=0.
\label{pressure10}
\end{equation}
In our case $B=C$ and, hence, $p_y=p_z$. We shall consider the case, where the tangential pressure $p_y=p_z=0$.
Now the equation (\ref{pressure10}) becomes 
\begin{equation}
p'+A(\rho+p) +2Bp = 0,
\label{pressure11}
\end{equation}
where $p\equiv p_x$. We have two unknown functions: $p$ and $A$. However, it is not convenient to try to find the relation between these functions using Eq. (\ref{pressure11}). It is better to take Eq. (\ref{Ein4n}) with the vanishing right-hand side:
\begin{equation}
A'+A^2+B'+B^2+AB=0.
\label{pressureless}
\end{equation}
The function $B$ still satisfies (\ref{Ein1n1}) and (\ref {B}); using (\ref {B}) we can rewrite (\ref {pressureless}) in terms of the function $a$:
\begin{equation}
a''-\frac23\tan k(x+x_0) a' +\left(\frac43k^2\tan^2 k(x+x_0) -\frac23\frac{k^2}{\cos^2 k(x+x_0)}\right)a = 0.
\label{pressureless1}
\end{equation}
Looking for the solution of these second order linear differential equation in the form 
\begin{equation}
a(x) (\cos k(x+x_0))^{\alpha}(\sin k(x+x_0))^{\beta}e^{k\gamma (x+x_0)},
\label{search}
\end{equation}
we find two sets of the parameters giving the solution of Eq. (\ref{pressureless1}):
\begin{eqnarray}
&&\alpha = -\frac13,\ \beta = 0,\ \gamma =\frac{1}{\sqrt{3}},\nonumber \\
&&\alpha = -\frac13,\ \beta = 0,\ \gamma =-\frac{1}{\sqrt{3}}.
\label{search1}
\end{eqnarray}
Thus, the general solution of Eq. (\ref{pressureless1}) is 
\begin{equation}
a(x) = (\cos k(x+x_0))^{-1/3}(a_4e^{\frac{1}{\sqrt{3}}k(x+x_0)}+a_5e^{-\frac{1}{\sqrt{3}}k(x+x_0)}).
\label{pless}
\end{equation}
Now, we find 
\begin{equation}
A = \frac{a'}{a} = \frac{k}{3}\tan k(x+x_0) + \frac{k}{\sqrt{3}}\frac{a_4e^{\frac{1}{\sqrt{3}}k(x+x_0)}-a_5e^{-\frac{1}{\sqrt{3}}k(x+x_0)}}
{a_4e^{\frac{1}{\sqrt{3}}k(x+x_0)}+a_5e^{-\frac{1}{\sqrt{3}}k(x+x_0)}}.
\label{pless1}
\end{equation}
Substituting this expression into Eq. (\ref{Ein2n}) we find the transversal pressure 
\begin{equation}
p = -\frac{4k^2}{3\sqrt{3}}\tan k(x+x_0)\frac{a_4e^{\frac{1}{\sqrt{3}}k(x+x_0)}-a_5e^{-\frac{1}{\sqrt{3}}k(x+x_0)}}
{a_4e^{\frac{1}{\sqrt{3}}k(x+x_0)}+a_5e^{-\frac{1}{\sqrt{3}}k(x+x_0)}}.
\label{pless2}
\end{equation}
In order to have the pressure vanishing at $x=-L$, we can again choose $x_0=L$. Then fixing 
\begin{equation}
a_5 = a_4e^{\frac{4kL}{\sqrt{3}}},
\label{pless3}
\end{equation} 
we have the pressure vanishing also at $x=L$. 
Finally, we have 
\begin{equation}
p = \frac{4k^2}{3\sqrt{3}}\tan k(x+L)\tanh \frac{k}{\sqrt{3}}(L-x),
\label{pless4}
\end{equation}
and 
\begin{equation}
a(x) = a_6(\cos k(x+L))^{-1/3}\cosh \frac{1}{\sqrt{3}}k(x-L).
\label{pless5}
\end{equation}
For $x > L$ this solution should be matched with the Weyl-Levi-Civita solution with the same value of the parameter $x_{WLC}$ as in the previous section. For $x < -L$ the obtained solution is matched with the Rindler solution with 
\begin{equation}
x_R=-L+\frac{\sqrt{3}\coth\frac{2kL}{\sqrt{3}}}{k}.
\label{pless6}
\end{equation}
It is easy to see that as long as $2kL<\pi/2$ the internal metric is regular and the pressure (\ref {pless4}) is finite everywhere in the slab. Thus, in contrast to the case of the Schwarzschild geometry, we have here a non-singular internal solution with an anisotropic pressure, namely with the pressure whose tangental components are identically equal to zero.

\section{Concluding remarks}
We have found two static solutions for an infinite slab of finite thickness immersed in the spacetime with plane symmetry. How are these solutions related to the solutions of a matter source localized on an infinitely thin plane? First of all let us note that our solutions are non-singular inside the slab if the condition (\ref{range}) is satisfied. If we introduce 
the notion of the energy of the unit square of the slab $M$:
\begin{equation}
M = 2\rho L = \frac{8k^2L}{3},
\label{M}
\end{equation}
then the condition (\ref{range}) becomes 
\begin{equation}
L < \frac{\pi^2}{12M}.
\label{range2}
\end{equation} 
Thus, if we fix the value of $M$ and begin squeezing the slab, diminishing $L$, we do not encounter anything similar to the Buchdahl limit for spherically symmetric configurations \cite{Buchdahl}.   In other words, if the relation (\ref{range2}) is satisfied at some value of $L_0$, it remain satisfied at all finite values of $L < L_0$. On the other hand, if we start increasing the thickness of the slab then at the value $L=  \frac{\pi^2}{12M}$ a singularity arises inside the slab. Moreover, in the case considered in the section 4 the pressure also becomes infinite. 

What happens when $L \rightarrow 0$? Obviously,  the energy density will tend to the delta function
\begin{equation}
\rho_{L\rightarrow 0} \rightarrow M\delta(x).
\label{M1}
\end{equation}
As was discussed in paper \cite{Fulling}, the tangential pressure should also tend to infinity to maintain the validity of the  energy-momentum conservation equation (\ref{pressure}). In our solution presented in Section 4, the tangential pressure is identically zero. One can show, using Eqs. (\ref{pressure2}) and (\ref{gen-sol1}), 
that in the solution with an isotropic pressure presented in Section 3, 
the pressure in the slab is limited by the value $p \approx M^2$ when $L \rightarrow 0$. Thus, while both of these solutions are well-defined at any arbitrary small, but finite 
value of the thickness parameter $L$, there is not a smooth transition to an infinitely thin plane configuration for these two solutions. However, these solutions represent 
some particular configurations acceptable from a physical point of view. 
Let us emphasise once again  that we did not fix some particular equation of state for the matter filling our slab. We simply required that the energy density on the slab is constant 
and that the pressure disappears at the boundaries of the slab. These conditions are the same used in the Schwarzschild internal solution \cite{Schwarz1}.  
Then we considered two particular additional conditions: one of them requires the isotropy of the pressure, just like in the Schwarzschild solution \cite{Schwarz1}, another requires the disappearance of the tangential pressure in all the slab. For both these requirements we have found exact solutions.  
In principle, one can imagine the existence of a solution where the transversal  and tangential pressures are different functions of the coordinate $x$, vanishing on the borders of the slab. Then, one cannot exclude that for some solutions of this kind a smooth transition 
to the localised matter configurations is possible.      
   
There is also another  problem here: it would be interesting to find matter distributions, which imply the existence of solutions of the Einstein equations which are matched in the empty regions of the space with the general spatial Kasner solutions   (\ref{Kasner}), (\ref{Kasner1}) with $p_2\neq p_3$. We hope to attack these problems in a future work \cite{we-future}.    
    
\section*{Acknowledgements}
 We are grateful to R. Casadio, J. Ovalle and G. Venturi for useful discussions.  
\\


\begin{thebibliography}{99}
\bibitem{Schwarz}
K.~Schwarzschild,
  Sitzungsber.\ Preuss.\ Akad.\ Wiss.\ Berlin (Math.\ Phys.\ ) {\bf 1916} (1916) 189.
\bibitem{Schwarz1}
K.~Schwarzschild,
  Sitzungsber.\ Preuss.\ Akad.\ Wiss.\ Berlin (Math.\ Phys.\ ) {\bf 1916} (1916) 424.
\bibitem{Tolman}
R.~C.~Tolman,
  Phys.\ Rev.\  {\bf 55} (1939) 364.
\bibitem{Op-Vol}
J.~R.~Oppenheimer and G.~M.~Volkoff,
  Phys.\ Rev.\  {\bf 55} (1939) 374.
\bibitem{Buchdahl}
H.~A.~Buchdahl,
  Phys.\ Rev.\  {\bf 116} (1959) 1027.
\bibitem{Fulling0}
C.~S.~Trendafilova and S.~A.~Fulling,
  Eur.\ J.\ Phys.\  {\bf 32} (2011) 1663
\bibitem{Geroch}
R.~P.~Geroch and J.~H.~Traschen,
  Phys.\ Rev.\ D {\bf 36} (1987) 1017.
\bibitem{Amundsen}
P.~A.~Amundsen and O.~Gron,
  Phys.\ Rev.\ D {\bf 27} (1983) 1731.
\bibitem{Fulling}
S.~A.~Fulling, J.~D.~Bouas and H.~B.~Carter,
  Phys.\ Scripta {\bf 90} (2015) no.8,  088006.
\bibitem{Kasner}
E.~Kasner,
  Am.\ J.\ Math.\  {\bf 43} (1921) 217.
\bibitem{Weyl}
H. Weyl, Annalen Physik {\bf 54} (1917) 117.
\bibitem{Levi-Civita}
T. Levi-Civita, Atti Accad. Naz. Rend. {\bf 27} (1918) 240.
\bibitem{Taub}
A.~H.~Taub,
  Annals Math.\  {\bf 53} (1951) 472.
\bibitem{Heck-Schuck}
O. Heckmann and E. Schucking, Handbuch der Physik {\bf 53} (1959) 489.
\bibitem{Khal-Lif}
E.~M.~Lifshitz and I.~M.~Khalatnikov,
  Adv.\ Phys.\  {\bf 12} (1963) 185.
\bibitem{BKL}
V.~A.~Belinsky, I.~M.~Khalatnikov and E.~M.~Lifshitz,
  Adv.\ Phys.\  {\bf 19} (1970) 525.
\bibitem{Misner}
C.~W.~Misner,
  Phys.\ Rev.\ Lett.\  {\bf 22} (1969) 1071.
\bibitem{Damour}
T.~Damour, M.~Henneaux and H.~Nicolai,
  Class.\ Quant.\ Grav.\  {\bf 20} (2003) R145.
\bibitem{Rindler}
W.~Rindler,
  Am.\ J.\ Phys.\  {\bf 34} (1966) 1174.
\bibitem{Shirkov}
N.~N.~Bogolyubov and D.~V.~Shirkov,
  Quantum Fields,
  Reading, Usa: Benjamin/cummings ( 1983).
\bibitem{Florides}
P. S. Florides, Proc. R. Soc. Lond. A {\bf 337} (1974) 529. 
\bibitem{we-future}
A.Yu. Kamenshchik and T. Vardanyan, work in progress.
\end{thebibliography}
\end{document}